\documentclass[letterpaper]{article} 
\pdfoutput=1
\usepackage{aaai24}  
\usepackage{times}  
\usepackage{helvet}  
\usepackage{courier}  
\usepackage[hyphens]{url}  
\usepackage{graphicx} 
\usepackage{natbib}  
\usepackage{caption} 
\usepackage{algorithm}
\usepackage{algorithmic}
\usepackage{amsmath}
\usepackage{amsfonts}
\usepackage{amssymb}
\usepackage{tabularx}
\usepackage{booktabs}
\usepackage{multirow}
\usepackage{newfloat}
\usepackage{listings}
\DeclareCaptionStyle{ruled}{labelfont=normalfont,labelsep=colon,strut=off} 
\floatstyle{ruled}
\newfloat{listing}{tb}{lst}{}
\floatname{listing}{Listing}
\title{Causally Aware Generative Adversarial Networks for Light Pollution Control}
\author {
    Yuyao Zhang,
    Ke Guo,
    Xiao Zhou\thanks{Corresponding Author.}
}
\affiliations {
    Gaoling School of Artificial Intelligence, Renmin University of China\\
    \{2020201710,2021201515,xiaozhou\}@ruc.edu.cn
}

\begin{document}

\maketitle

\begin{abstract}
Artificial light plays an integral role in modern cities, significantly enhancing human productivity and the efficiency of civilization. However, excessive illumination can lead to light pollution, posing non-negligible threats to economic burdens, ecosystems, and human health. Despite its critical importance, the exploration of its causes remains relatively limited within the field of artificial intelligence, leaving an incomplete understanding of the factors contributing to light pollution and sustainable illumination planning distant. To address this gap, we introduce a novel framework named \underline{C}ausally \underline{A}ware \underline{G}enerative \underline{A}dversarial \underline{N}etworks (CAGAN). This innovative approach aims to uncover the fundamental drivers of light pollution within cities and offer intelligent solutions for optimal illumination resource allocation in the context of sustainable urban development. We commence by examining light pollution across 33,593 residential areas in seven global metropolises. Our findings reveal substantial influences on light pollution levels from various building types, notably grasslands, commercial centers and residential buildings as significant contributors. These discovered causal relationships are seamlessly integrated into the generative modeling framework, guiding the process of generating light pollution maps for diverse residential areas. Extensive experiments showcase CAGAN’s potential to inform and guide the implementation of effective strategies to mitigate light pollution. Our code and data are publicly available at https://github.com/zhangyuuao/Light\_Pollution\_CAGAN.
\end{abstract}

\section{Introduction}

When Thomas Edison's electric light first illuminated a street in New York City, it marked the dawn of a modern era intertwined with artificial illumination \cite{doi:10.1289/ehp.117-a20}. 
Since that moment, our world has become saturated with artificial light, transforming environments and enhancing economic productivity. However, the seemingly benign nature of artificial light conceals substantial adverse effects on human health. Nocturnal exposure to artificial light suppresses the synthesis and secretion of pineal melatonin, a hormone crucial for regulating the circadian rhythm \cite{FALCHI20112714,CAO2023589,Leena2019Systematic}, leading to an increased risk of metabolic disorders, cardiovascular issues, and heart disease \cite{brugger1995impaired, FALCHI20112714, CAO2023589}. Considering the widely acknowledged reality that urban areas serve as primary generators of artificial light, affecting both the adjacent sky and ground~\cite{doi:10.1177/0885412220986421}, coupled with the rapid acceleration of urbanization leading more individuals to inhabit these landscapes, there is a compelling need to scrutinize the intricate implications of artificial light. This underscores the importance of managing and curbing light pollution, presenting a crucial challenge for the sustainable development of contemporary societies worldwide.


To uncover and probe the phenomenon of light pollution within urban areas, several prior studies have delved into this domain. Notably, some investigations utilized satellite-based night-time light (NTL) imagery to quantify and evaluate the extent of light pollution~\cite{zheng2021africa,mu2021evaluation,cox2020national,hu2020global}. Others employed a combination of NTL imagery and Geographic Information System (GIS) techniques to analyze the intricate interplay of light pollution~\cite{butt2012estimation,chalkias2006modelling}. Moreover, the distribution of Points of Interest (POIs) in urban settings significantly shapes the occurrence of light pollution. As a valuable data source in urban computing, POIs play a crucial role in tasks such as discovering new venues~\cite{zhou2019topic}, predicting chronic disease rates~\cite{wang2018predicting}, allocating resources~\cite{zhou2018discovering}, and evaluating investments' impact on socio-economic indicators~\cite{zhou2017cultural}. Of particular note, a recent study used NTL imagery and POI data to comprehensively assess citywide light pollution~\cite{zhao2021assessing}.



However, existing studies on light pollution exhibit specific limitations. The majority of research tends to superficially quantify light pollution levels in urban areas, often neglecting the underlying causes. Consequently, practical insights for effective real-world policy development are scarce. Additionally, these investigations commonly concentrate on entire cities, overlooking nuanced analyses at smaller spatial scales or specific land use functions \cite{butt2012estimation,chalkias2006modelling,zhao2021assessing}. Furthermore, the evaluation metrics employed in these studies are frequently uniform and rudimentary, disregarding the complexities of localized circumstances. For instance, equating illumination levels across different areas with the degree of light pollution is illogical, as certain regions may legitimately require higher luminance without causing significant impacts. In summary, the current literature lacks a comprehensive framework that holistically addresses urban light pollution, spanning detection, assessment, and suggestion formulation, while considering solid causal relationships at finely-grained spatial and functional scales.


To bridge this gap, we introduce a novel framework, Causally Aware Generative Adversarial Networks (CAGAN), meticulously crafted to unveil causal relationships between urban areas and light pollution. These relationships subsequently steer the process of generating light pollution maps across diverse residential regions. To provide a comprehensive and objective assessment of light pollution levels, we incorporate three indices: \textit{over illumination}~\cite{mishra2018photoperiodic}, \textit{light trespass}~\cite{schreuder1986light}, and \textit{light clutter} ~\cite{rajkhowa2014light}. By holistically considering these indices, we effectively gauge the intricate impact of light pollution on residential areas. Furthermore, our methodology includes a thorough exploration employing causal inference techniques, which unveil significant effects of different building types on light pollution levels. These identified causal relationships are seamlessly integrated into CAGAN, thereby enhancing the precision of generated light pollution maps in urban areas. We assess light pollution across seven globally significant metropolises, renowned for their robust economies and international importance. Rigorous experiments underscore CAGAN's capacity to provide valuable insights and aid in the development and implementation of strategies to mitigate the adverse impacts of light pollution.

Our main contributions are summarized as follows:
\begin{itemize}
    
    \item We propose an evaluation framework for light pollution that enables comprehensive and objective assessment of light pollution extent in various urban residential areas.
    
    \item We investigate the underlying factors contributing to light pollution across metropolises, unveiling causal relationships through advanced causal inference techniques.

    \item We utilize the identified causal relationships related to light pollution as conditional information to guide the process of generating light pollution maps, which empowers local administrators to make informed decisions and allocate lighting resources effectively.
\end{itemize}

\section{Related Work}
In this section, we explore three categories of studies closely related to our topic and methodology. These include research on light pollution assessment, applications of causal inference, and the use of Generative Adversarial Networks (GANs) in the urban computing domain.

\subsubsection{Light Pollution Assessment.}


The utilization of nighttime light images to evaluate light pollution levels has garnered considerable attention in research. Notably, \citeauthor{zhao2021assessing} \shortcite{zhao2021assessing} enhanced urban-scale modeling and analysis by integrating POI data into their methodology. Additionally, \citeauthor{TONG2022155681}\shortcite{TONG2022155681} and \citeauthor{BURT2023}\shortcite{BURT2023} introduced distinctive indices for assessing light pollution. The ALAN Lab at the University of Hong Kong has conducted extensive academic investigations into local urban light pollution \cite{hkulightpollution}. In contrast, our research zeroes in on the impact of light pollution on residents in the residential areas of metropolises. We conduct a comprehensive assessment of light pollution levels in these areas, leveraging nighttime light images and POI data.

\subsubsection{Causal Inference in Urban Computing.}


The exploration of causal relationships within the realm of social science is crucial, and various methodologies in causal inference have yielded substantial insights in urban computing and analysis \cite{BAUMSNOW20153}. Overcoming confounding factors lies at the heart of causal inference, particularly in scenarios with high-dimensional confounders. To address this challenge, Propensity and Propensity Score Matching were introduced by \citeauthor{propensity} \shortcite{propensity}. Additionally, \citeauthor{morgan2015counterfactuals} \shortcite{morgan2015counterfactuals} proposed the Inverse of Propensity Weighting as an effective solution to confounder interference. The application of the causal inference framework by \citeauthor{10.1145/3494990} \shortcite{10.1145/3494990} in individual healthy lifestyle and mobility-related health policymaking further underscores its efficacy. In this study, we employ the Debiased Machine Learning method, as proposed by \citeauthor{chetverikov2016double} \shortcite{chetverikov2016double}. This method focuses on diminishing the influence of confounding variables, integrating machine learning models, and providing confidence intervals to enhance stability and effectiveness. 

\subsubsection{Application of GANs in Urban Computing.}


Researchers have harnessed the potential of GANs to create diverse urban environments at various scales. CityGAN, for instance, focuses on synthesizing architectural features and building images \cite{bachl2020citygan}. SG-GAN, introduced by \citeauthor{li2018semanticaware} \shortcite{li2018semanticaware}, enhances semantic segmentation on Cityscapes using virtual data. Domain-adaptive networks based on CycleGAN, proposed by \citeauthor{GUO2020127} \shortcite{GUO2020127}, generate urban scenes from virtual video games to aid in segmentation. In a different way, \citeauthor{9205243} \shortcite{9205243} employed conditional GANs to synthesize satellite-like urban images from historical maps. The work of \citeauthor{8518032} \shortcite{8518032} delves into simulating hyper-realistic urban patterns through GANs. \citeauthor{MetroGAN} \shortcite{MetroGAN} developed MetroGAN to utilize NTL images, water area data, and built-up area data to simulate urban morphology. However, our endeavor transcends the task of generating urban visuals or conducting style transfers for urban scenes. We leverage the capabilities of conditional variational auto-encoders to embed the contextual and semantic intricacies associated with light pollution into the generated maps. This infusion of information grants the maps interpretability, elucidating the implications of light pollution's effects.

\section{Preliminaries}


In this section, we outline methodologies for data collection, processing, index calculation, and constructing an evaluation system for residential light pollution.

\subsection{Data Collection and Processing}


Before initiating our study, we first selected seven global metropolises that stand out for their scale of urbanization and extent of economic development. These metropolises include New York, Los Angeles, Paris, London, Shanghai, Beijing, and Guangzhou. We established latitude and longitude boundaries for each metropolis and employed OpenStreetMap\footnote{https://openstreetmap.org}'s API to retrieve POI data spanning various land use types. Subsequently, we sampled the corresponding nighttime light intensity value from the remote sensing image of night-time light \cite{DVN/YGIVCD_2020} using the longitude and latitude coordinates of each POI. 
Taking into account the quantity of each land use type and the distribution of night-time light intensity values, we distilled a total of nine distinct POI categories in this study. These encompass brownfield, commercial, construction, farmland, forest, grass, industrial, residential, and retail.

\subsection{Light Pollution Evaluation}

\begin{figure}[htbp]
  \centering
  \includegraphics[width=\linewidth]{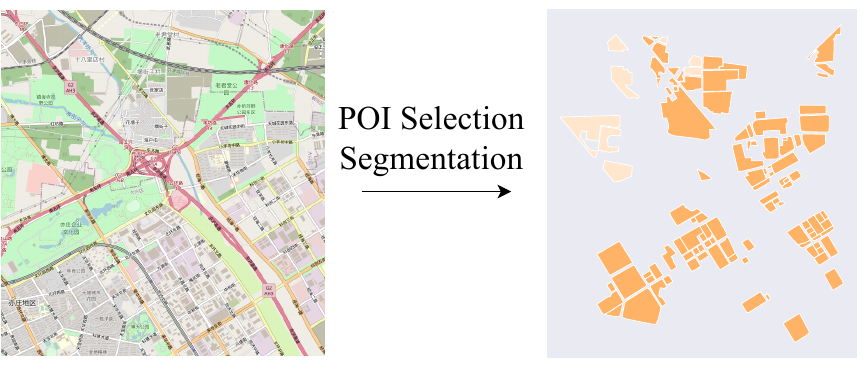}
  \caption{The segmentation of POI in a residential area.}
  \label{seg}
\end{figure}

Following the sampling process, we characterize a residential area within the metropolis as a square region centered around the "residential" land use type POI. 
This designated residential area is intended to encompass a diverse range of buildings that cater to a variety of residents' needs, with each occupying an approximate area of four square kilometers. Subsequently, we identify the POI situated within this residential zone and partition them into distinct plots, as depicted in Figure \ref{seg}.

\begin{figure}[htbp]
  \centering
  \includegraphics[width=\linewidth]{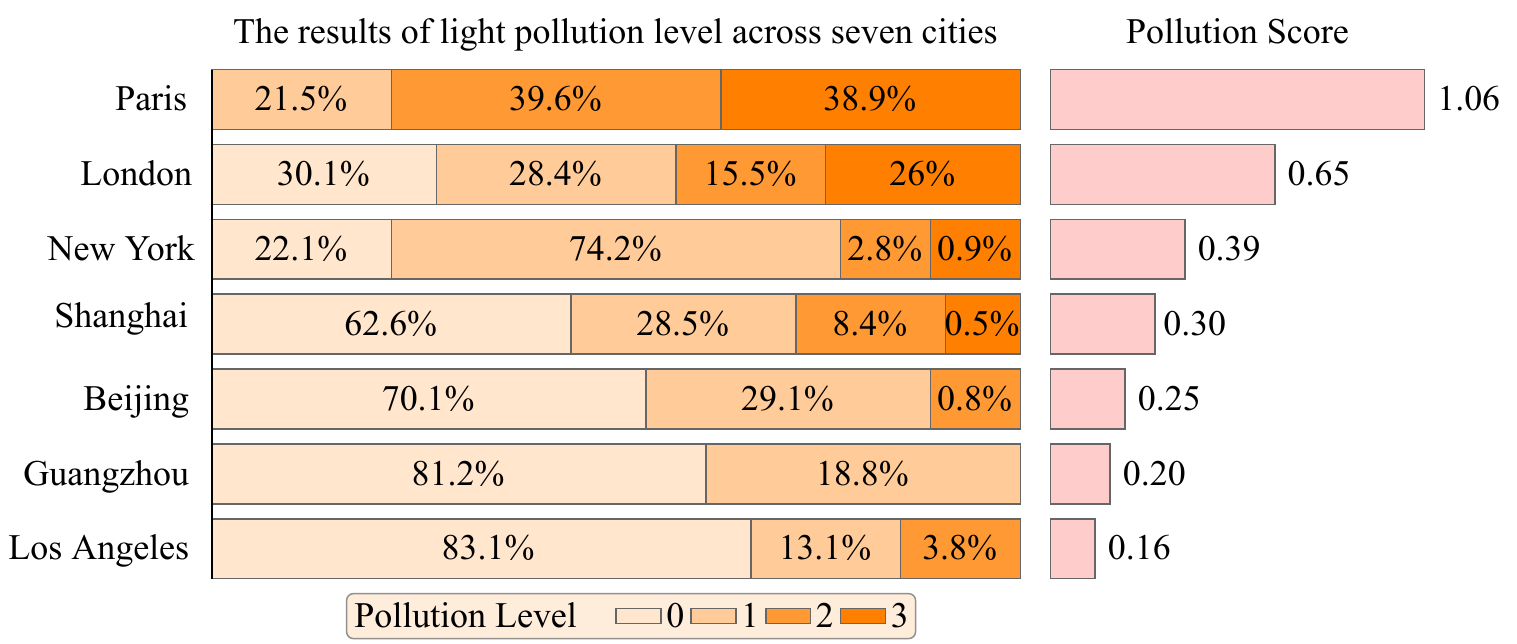}
  \caption{The results of light pollution assessment in all residential areas across the seven metropolises.}
  \label{result1}
\end{figure}

\begin{figure*}[t]
  \centering
  \includegraphics[width=.97\linewidth]{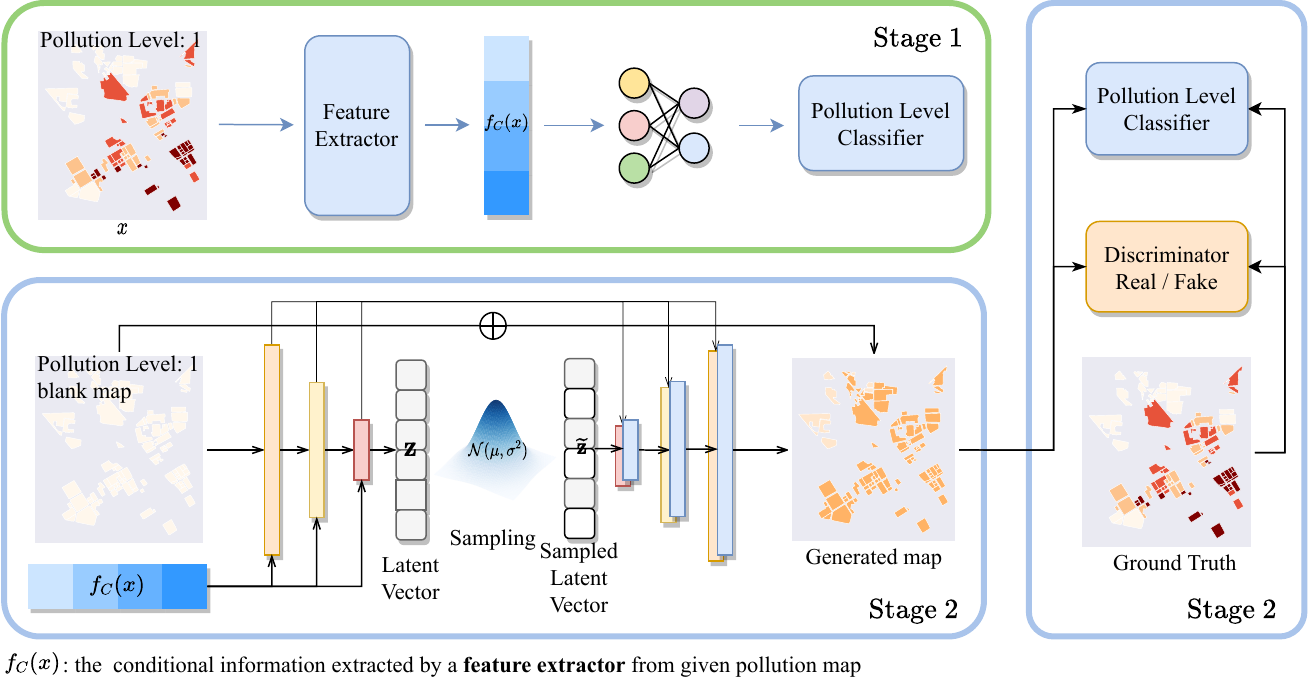}
  \caption{The comprehensive architecture of the proposed CAGAN, containing two training stages.}
  \label{CAGAN}
\end{figure*}

In assessing the impact of light pollution sources near residential areas, we drew inspiration from prior research that employed radial basis functions to gauge the extent of light pollution influence~\cite{zhao2021assessing}. This influence is quantified using the following formula:

\begin{equation}
    I^i_j = \mathrm{NTL}_j \times \frac{1}{\sqrt{2\pi}D} \times \exp{\left(-\frac{d^{i2}_j}{2D^2}\right)}
\end{equation}

\noindent where $I^i_j$ represents the level of light pollution impact on residential area $i$'th from $j$'th pollution source, $\mathrm{NTL}_j$ represents the night-time light intensity of $j$'th pollution source, $d^i_j$ represents the distance between the light pollution source and the residential area, and $D$ is the bandwidth in the radial basis function, which is a hyperparameter with a value of 1500m in this study.

We proceed to employ \textit{Total Nighttime Light} (TNL), \textit{Nighttime Light Disturbance} (NLD), and \textit{Nighttime Light Standard Deviation} (NLSD) to quantify the three indices, namely \textit{over illumination}, \textit{light trespass}, and \textit{light clutter}, as previously discussed~\cite{BURT2023,TONG2022155681}. In this context, let $r_i$ represent the $i$-th residential area, and $p_j$ symbolize the $j$-th light pollution source, the corresponding light pollution scores can be obtained through:

\begin{align}
    \mathrm{TNL}_i &= \sum_{p_j \in \mathcal{C}(r_i)} I^i_j \\
    \mathrm{NLD}_i &= \sum_{p_j \in \mathcal{C}(r_i)} I^i_j - I^i_i \\
    \mathrm{NLSD}_i &= \sqrt{\frac{\sum_{p_j \in \mathcal{C}(r_i)}(I^i_j - \mu^i_j)^2}{|\mathcal{C}(r_i)|}} \\
    \mathrm{Score}_i &= \mathrm{TNL}_i + \mathrm{NLD}_i + \mathrm{NLSD}_i
\end{align}

\noindent where $\mathcal{C}(r_i)$ represents the residential area centered on $r_i$. 

By computing these three light pollution indices, we can assess the influence of light pollution on residential areas from varied perspectives. Subsequently, we employ unsupervised learning algorithms to categorize these areas into five distinct levels of light pollution, discerning which residential regions are prone to enduring significant light pollution and which areas enjoy better conditions. Armed with this insight, we can offer recommendations for addressing severe light pollution in residential zones. This guidance aids local residents in enhancing the rational distribution of light and managing pollution levels around nearby POIs.

We meticulously calculate light pollution indicators for all residential areas within the seven metropolises. Subsequently, we choose a suitable number of centroids and apply the k-means clustering algorithm to divide light pollution levels into four distinct categories. The classification of residential areas in each metropolis, based on their pollution levels, is illustrated in Figure \ref{result1}. Additionally, the light pollution scores for each city are displayed on the right side of the figure, facilitating a visual comparison of residential light pollution at the urban scale.


\section{Methodology}
In this section, we introduce the Causally Aware Generative Adversarial Networks (CAGAN) framework for generating fine-grained light pollution maps in urban environments while considering causal relationships. The framework is divided into two main phases: 1) estimating the causal effects of light pollution through debiased machine learning, and 2) producing coherent light pollution maps by integrating the obtained conditional causal information into a novel GAN framework. Next, we delve into a comprehensive explanation of the techniques employed in each phase.

\subsection{Causal Inference}

Causal inference is a scientific method crucial for identifying causal relationships among events or variables. It allows us to determine if one event truly causes another, moving beyond simple correlations or coincidences, by examining patterns and connections in observed phenomena. In our study, applying causal inference helps us investigate the complex links between residential light pollution and various POIs in a specific area, uncovering fundamental causal mechanisms. Through this approach, we make significant contributions to improving and managing light pollution.


We incorporate the causal inference within the latent outcome framework, employing a two-stage Debiased Machine Learning (DML) approach as the cornerstone of our research methodology. DML can be used to estimate treatment effects, particularly in scenarios where all potential confounding variables that exert a direct influence on both the treatment decision and observed outcome are accessible~\cite{chetverikov2016double}. To elaborate, let $Y$ denote the outcome variable encompassing three light pollution indices for each residential area. For each instance, we meticulously designate a specific category of POIs and calculate its quantity, average nighttime light intensity, and mean distance from the center of the residential area, all of which are classified as intervention variables labeled as $T$. Additionally, we account for other variables connected to POIs, denoted as $X$, which serve as potential confounders.

\begin{figure}[h]
  \centering
  \includegraphics[width=\linewidth]{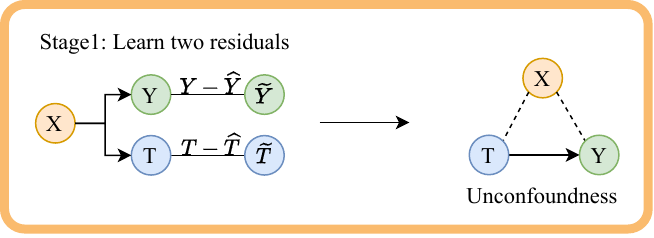}
  \caption{Regression from confounders $X$ to treatment $T$ and outcome $Y$ to achieve unconfoundness.}
  \label{4}
\end{figure}

The DML simplifies the challenge by initially focusing on estimating two predictive tasks (see Figure \ref{4}):

\begin{itemize}
    \item[1.] Predicting the outcome $Y$ from the confounders $X$
    \begin{equation}
    \widehat{Y} = f(X) + \epsilon, \ \mathbb{E}(\epsilon | X) = 0
    \end{equation}
    \item[2.] Predicting the treatment $T$ from the confounders $X$
    \begin{equation}
    \widehat{T} = g(X) + \eta, \ \mathbb{E}(\eta | X) = 0
    \end{equation}
\end{itemize}
\noindent Furthermore, it holds that $\mathbb{E}(\epsilon \cdot \eta | X) = 0$.

\begin{figure}[h]
  \centering
  \includegraphics[width=\linewidth]{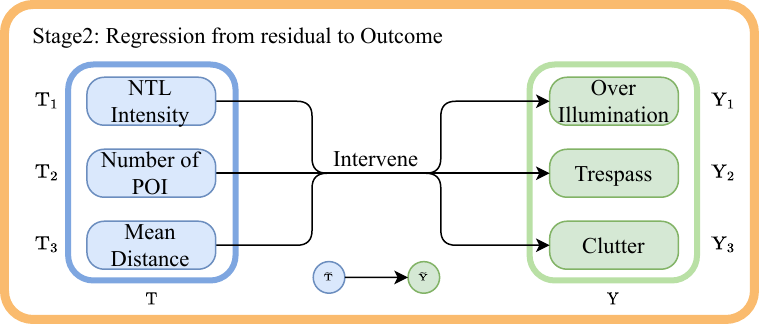}
  \caption{Estimation of Average Treatment Effect.}
  \label{5}
\end{figure}

Subsequently, we combine these two residuals using predictive models during the final stage of estimation, thereby constructing a model for the Average Treatment Effect (ATE), as illustrated in Figure \ref{5}. Assuming that the ultimate goal is to estimate $\theta$, the equation is as follows:
\begin{align}
    &\widetilde{Y} = Y - f(X) \\
    &\widetilde{T} = T - g(X) = \eta \\
    &\widetilde{Y} = \theta(X)\widetilde{T} + \epsilon
\end{align}

Given that $\mathbb{E}(\epsilon \cdot \eta | X) = 0$, our objective is to estimate $\theta(X)$ along with the optimal parameters, representing the desired ATE:
\begin{equation}
    \theta^* = \arg\min_{\theta} \mathbb{E}[(\widetilde{Y}-\theta(X)\widetilde{T})^2]
\end{equation}

The novelty of DML resides in the orthogonal relationship between the hyperplane formed by confounders and residuals. Through the regression of the two residuals, we effectively mitigate the confounding bias originating from the confounders on the treatment variables. Thus, we attain an unbiased estimation of the causal effect of the treatment.

\subsection{Causally Aware Generative Adversarial Networks}

The assessment of authenticity in generated images often concerns resolution and detail. However, for city light pollution maps, this criteria emphasizing high resolution and details may not be crucial. In this context, the emphasis shifts towards generating distinct blocks with clear boundaries.


The proposed Causally Aware Generative Adversarial Networks (CAGAN) comprises two distinct training stages, as illustrated in Figure \ref{CAGAN}. In the initial stage, a light pollution image classifier is trained with actual light pollution images, enabling it to establish accurate associations with given residential area light pollution maps. Subsequently, a conditional variational autoencoder is trained exclusively using the base map without colorization. The generator is then guided by the classifier and discriminator to progressively generate authentic light pollution maps. Upon completion of both stages, the model is equipped to generate accurate light pollution maps for residential areas based on the relevant conditional information.

Specifically, our variational inference module employs the ELBO (Evidence Lower Bound) loss function, which incorporates reconstruction error based on L1 norm and KL divergence error.  The reconstruction error measures the discrepancy between generated images and the original input images. The KL divergence error is used to assess the difference between the generated latent variable distribution and the prior latent variable distribution. By minimizing the divergence between the generated images and the prior distribution, the generative model learns meaningful latent representations. To be specific,
\begin{align}
    \mathcal{L}_{KL} &= \frac{1}{2}\left(\boldsymbol{\mu}^{\top} \boldsymbol{\mu}+\operatorname{sum}(\exp (\boldsymbol{\epsilon})-\boldsymbol{\epsilon}-1)\right) \\
    \mathcal{L}_{G} &= \Vert\boldsymbol{x}- \widehat{\boldsymbol{x}}\Vert_1 \\
    \mathcal{L}_{ELBO} &= \lambda_1 \mathcal{L}_{G} + \lambda_2 \mathcal{L}_{KL} 
\end{align}
\noindent where $\boldsymbol{\mu}$ and $\boldsymbol{\epsilon}$ are mean and covariance of the latent vector output by encoder, and $\boldsymbol{x}$ is the original light pollution map.

Furthermore, our approach incorporates the following loss functions: cross-entropy loss, which measures the disparity between the predicted pollution level and the true labels during image classification; binary cross-entropy (BCE) loss, employed to discern the authenticity of the generated maps; and mean squared error (MSE), assessing the difference between the extracted features $f_C(x)$ of the generated and real maps, as outlined in CVAE-GAN \cite{DBLP:conf/iccv/BaoCWLH17}.
\begin{align}
    \mathcal{L}_C &= -\sum_{i=1}^{N} \sum_{c=1}^{C} y_{i,c} \log(p_{i,c}) \\
    \mathcal{L}_D &= -\sum_{i=1}^{N} [y_i \log(p_i) + (1-y_i) \log(1-p_i)] \\
    \mathcal{L}_{G D} &= \left\|\mathbb{E}_{\boldsymbol{x} \sim P_{r}} f_{D}(\boldsymbol{x})-\mathbb{E}_{\widehat{\boldsymbol{x}} \sim P_{\widehat{\boldsymbol{x}}}} f_{D}(\widehat{\boldsymbol{x}})\right\|_{2}^{2} \\
    \mathcal{L}_{G} &= \left\|f_{D}(\boldsymbol{x})-f_{D}\left(\widehat{\boldsymbol{x}}\right)\right\|_{2}^{2} + \left\|f_{C}(\boldsymbol{x})-f_{C}\left(\widehat{\boldsymbol{x}}\right)\right\|_{2}^{2}
\end{align}
\noindent Here, $N$ represents the number of training samples, $C$ denotes the number of pollution levels, $y_{i,c}$ is the classification label, $y_i$ denotes the label for true or fake maps, and $f_D$ and $f_C$ represent the features extracted from the input map $\boldsymbol{x}$.

Ultimately, the complete loss function for optimization is:
\begin{equation}
    \mathcal{L} = \mathcal{L}_{ELBO} + \lambda_3\mathcal{L}_{C} + \lambda_4\mathcal{L}_{D} + \lambda_5\mathcal{L}_{G D} + \lambda_6\mathcal{L}_{G}
\end{equation}
\noindent


\section{Experiments}


\subsection{Experimental Setup}

\subsubsection{Data Processing.} After addressing the outliers, we divide the dataset into training, validation, and test sets in a 0.7:0.15:0.15 ratio, yielding 23,514 valid training samples. Additionally, we set various random number seeds to generate different data subsets, and subsequently test and confirm the effectiveness of the training.


\subsubsection{Causal Inference Experiment Setup.} We use a multi-task elastic net in linear models to efficiently fit two residuals. Additionally, we employ a low-dimensional linear regression for estimating the Average Treatment Effect (ATE). Our methodology involves the use of 3-fold cross-validation, setting the maximum number of iterations at 2000.

\subsubsection{CAGAN Experiment Setup.} 

In the initial stage, a classifier is trained using ground truth light pollution maps with pollution levels as labels. Employing a pre-trained ResNet50 as the backbone network, augmented by a fully connected layer for regressing conditional vectors, we utilize a classification head and the conditional vector for pollution level classification. Subsequently, in the second stage, the classifier's parameters are kept fixed, and training proceeds for the generator and discriminator networks. The models are trained using the methodology-specific loss functions. For our experiments, we set hyperparameters as follows: $\lambda_1 = 1, \lambda_2 = 0.1, \lambda_3 = 1, \lambda_4 = 1, \lambda_5 = 0.001, \lambda_6 = 1$.



\subsection{Experimental Results}
\begin{table*}[h!]
\centering
\resizebox{\linewidth}{!}{
    \begin{tabular}{cc|ccc|ccc|ccc}
    \toprule
    \fontsize{21}{7}\selectfont Indices & \fontsize{21}{7}\selectfont City & \fontsize{21}{7}\selectfont Residential  & \fontsize{21}{7}\selectfont Commercial  & \fontsize{21}{7}\selectfont Retail & \fontsize{21}{7}\selectfont Brownfield  & \fontsize{21}{7}\selectfont Construction & \fontsize{21}{7}\selectfont Industrial & \fontsize{21}{7}\selectfont Grass & \fontsize{21}{7}\selectfont Farmland & \fontsize{21}{7}\selectfont Forest \\ \midrule
     \multirow{7}{*}{\fontsize{21}{7}\selectfont Illumination}
    & \fontsize{21}{7}\selectfont Paris & \fontsize{21}{7}\selectfont $0.318^{***}$ & \fontsize{21}{7}\selectfont $0.004$ & \fontsize{21}{7}\selectfont $0.016$ & \fontsize{21}{7}\selectfont $-0.089^{*}$ & \fontsize{21}{7}\selectfont $0.165^{**}$ & \fontsize{21}{7}\selectfont $-0.202^{***}$ & \fontsize{21}{7}\selectfont $1.414^{***}$ & \fontsize{21}{7}\selectfont $-0.026$ & \fontsize{21}{7}\selectfont $0.303^{**}$ \\
    
    & \fontsize{21}{7}\selectfont London & \fontsize{21}{7}\selectfont $0.243^{***}$ & \fontsize{21}{7}\selectfont $0.279^{***}$ & \fontsize{21}{7}\selectfont $0.350^{***}$ & \fontsize{21}{7}\selectfont $0.001$ & \fontsize{21}{7}\selectfont $0.294^{***}$ & \fontsize{21}{7}\selectfont $0.038^{***}$ & \fontsize{21}{7}\selectfont $0.140^{***}$ & \fontsize{21}{7}\selectfont $0.175^{***}$ & \fontsize{21}{7}\selectfont $0.238^{***}$ \\

    & \fontsize{21}{7}\selectfont Newyork & \fontsize{21}{7}\selectfont $0.312^{***}$ & \fontsize{21}{7}\selectfont $0.144^{***}$ & \fontsize{21}{7}\selectfont $-0.009$ & \fontsize{21}{7}\selectfont $0.093^{**}$ & \fontsize{21}{7}\selectfont $0.150^{***}$ & \fontsize{21}{7}\selectfont $0.008$ & \fontsize{21}{7}\selectfont $0.906^{***}$ & \fontsize{21}{7}\selectfont $-0.147^{***}$ & \fontsize{21}{7}\selectfont $0.082^{***}$ \\
    
    & \fontsize{21}{7}\selectfont Shanghai & \fontsize{21}{7}\selectfont $0.422^{***}$ & \fontsize{21}{7}\selectfont $0.279^{***}$ & \fontsize{21}{7}\selectfont $0.184^{***}$ & \fontsize{21}{7}\selectfont $0.083^{***}$ & \fontsize{21}{7}\selectfont $0.065^{***}$ & \fontsize{21}{7}\selectfont $0.160^{***}$ & \fontsize{21}{7}\selectfont $0.573^{***}$ & \fontsize{21}{7}\selectfont $0.007$ & \fontsize{21}{7}\selectfont $0.298^{***}$ \\
    
    & \fontsize{21}{7}\selectfont Beijing & \fontsize{21}{7}\selectfont $0.301^{***}$ & \fontsize{21}{7}\selectfont $0.202^{***}$ & \fontsize{21}{7}\selectfont $0.120^{***}$ & \fontsize{21}{7}\selectfont $0.128^{***}$ & \fontsize{21}{7}\selectfont  $0.133^{***}$ & \fontsize{21}{7}\selectfont $0.133^{***}$ & \fontsize{21}{7}\selectfont $0.681^{***}$ & \fontsize{21}{7}\selectfont
    $0.021$ & \fontsize{21}{7}\selectfont $0.023$ \\
    
    & \fontsize{21}{7}\selectfont Guangzhou & \fontsize{21}{7}\selectfont $0.518^{***}$ & \fontsize{21}{7}\selectfont
     $0.421^{***}$ & \fontsize{21}{7}\selectfont $0.130^{***}$ & \fontsize{21}{7}\selectfont $0.225^{***}$ & \fontsize{21}{7}\selectfont
      $0.332^{***}$ & \fontsize{21}{7}\selectfont $0.135^{***}$ & \fontsize{21}{7}\selectfont $0.638^{***}$ & \fontsize{21}{7}\selectfont $0.045$ & \fontsize{21}{7}\selectfont $0.182^{***}$\\
    
    & \fontsize{21}{7}\selectfont Los Angeles & \fontsize{21}{7}\selectfont $0.417^{***}$ & \fontsize{21}{7}\selectfont $0.145^{***}$ & \fontsize{21}{7}\selectfont $0.112^{***}$ & \fontsize{21}{7}\selectfont $0.108^{***}$ & \fontsize{21}{7}\selectfont $0.052^{***}$ & \fontsize{21}{7}\selectfont $0.051^{***}$ & \fontsize{21}{7}\selectfont $0.913^{***}$ & \fontsize{21}{7}\selectfont $-0.028$ & \fontsize{21}{7}\selectfont $0.028^{***}$ \\ \midrule

    \multirow{7}{*}{\fontsize{21}{7}\selectfont Trespass}

    & \fontsize{21}{7}\selectfont Paris & \fontsize{21}{7}\selectfont $0.195^{**}$ & \fontsize{21}{7}\selectfont $-0.019$ & \fontsize{21}{7}\selectfont $-0.024$ & \fontsize{21}{7}\selectfont $-0.139^{**}$ & \fontsize{21}{7}\selectfont $0.187^{**}$ & \fontsize{21}{7}\selectfont $0.057$ & \fontsize{21}{7}\selectfont $ 0.959^{***}$ & \fontsize{21}{7}\selectfont $-0.012$ & \fontsize{21}{7}\selectfont $0.325^{*}$ \\
    
    & \fontsize{21}{7}\selectfont London & \fontsize{21}{7}\selectfont $0.311^{***}$ & \fontsize{21}{7}\selectfont $0.208^{***}$ & \fontsize{21}{7}\selectfont $0.294^{***}$ & \fontsize{21}{7}\selectfont $-0.009$ & \fontsize{21}{7}\selectfont $0.266^{***}$ & \fontsize{21}{7}\selectfont $0.024^{*}$ & \fontsize{21}{7}\selectfont $0.057^{***}$ & \fontsize{21}{7}\selectfont $0.179^{***}$ & \fontsize{21}{7}\selectfont $0.244^{***}$ \\
    
    & \fontsize{21}{7}\selectfont Newyork & \fontsize{21}{7}\selectfont $0.214^{***}$ & \fontsize{21}{7}\selectfont $0.084^{***}$ & \fontsize{21}{7}\selectfont $-0.051^{***}$ & \fontsize{21}{7}\selectfont $0.037$ & \fontsize{21}{7}\selectfont $0.121^{***}$ & \fontsize{21}{7}\selectfont $-0.015$ & \fontsize{21}{7}\selectfont $0.628^{***}$ & \fontsize{21}{7}\selectfont $-0.155^{***}$ & \fontsize{21}{7}\selectfont $0.021$ \\
    
    & \fontsize{21}{7}\selectfont Shanghai & \fontsize{21}{7}\selectfont $0.380^{***}$ & \fontsize{21}{7}\selectfont $0.267^{***}$ & \fontsize{21}{7}\selectfont $0.147^{***}$ & \fontsize{21}{7}\selectfont $0.085^{***}$ & \fontsize{21}{7}\selectfont $0.043^{***}$ & \fontsize{21}{7}\selectfont $0.125^{***}$ & \fontsize{21}{7}\selectfont $0.431^{***}$  & \fontsize{21}{7}\selectfont $0.017^{**}$ & \fontsize{21}{7}\selectfont $0.247^{***}$  \\
    
    & \fontsize{21}{7}\selectfont Beijing & \fontsize{21}{7}\selectfont $0.107^{***}$ & \fontsize{21}{7}\selectfont $0.066^{***}$ & \fontsize{21}{7}\selectfont $0.075^{***}$ & \fontsize{21}{7}\selectfont $0.071^{***}$ & \fontsize{21}{7}\selectfont
    $0.095^{***}$ & \fontsize{21}{7}\selectfont $0.098^{***}$ & \fontsize{21}{7}\selectfont $0.360^{***}$ & \fontsize{21}{7}\selectfont $-0.003$ & \fontsize{21}{7}\selectfont $0.008$ \\
    
    & \fontsize{21}{7}\selectfont Guangzhou & \fontsize{21}{7}\selectfont $0.414^{***}$  & \fontsize{21}{7}\selectfont $0.265^{***}$ & \fontsize{21}{7}\selectfont $0.055^{***}$ & \fontsize{21}{7}\selectfont $0.211^{***}$ & \fontsize{21}{7}\selectfont
     $0.170^{***}$ & \fontsize{21}{7}\selectfont $0.114^{***}$ & \fontsize{21}{7}\selectfont $0.458^{***}$ & \fontsize{21}{7}\selectfont
      $0.038^{*}$ & \fontsize{21}{7}\selectfont $0.077^{***}$ \\
    
    & \fontsize{21}{7}\selectfont Los Angeles & \fontsize{21}{7}\selectfont $0.020$ & \fontsize{21}{7}\selectfont $0.330^{***}$ & \fontsize{21}{7}\selectfont $0.072^{**}$ & \fontsize{21}{7}\selectfont $0.182^{***}$ & \fontsize{21}{7}\selectfont $-0.321^{***}$ & \fontsize{21}{7}\selectfont $-0.148^{***}$ & \fontsize{21}{7}\selectfont $-0.064^{***}$ & \fontsize{21}{7}\selectfont $-0.000$ & \fontsize{21}{7}\selectfont $0.032$ \\ \midrule

    \multirow{7}{*}{\fontsize{21}{7}\selectfont Clutter}

     & \fontsize{21}{7}\selectfont Paris & \fontsize{21}{7}\selectfont $0.310^{**}$ & \fontsize{21}{7}\selectfont $0.110$ & \fontsize{21}{7}\selectfont $0.137$ & \fontsize{21}{7}\selectfont $-0.018$ & \fontsize{21}{7}\selectfont $-0.027$ & \fontsize{21}{7}\selectfont $0.243^{**}$ & \fontsize{21}{7}\selectfont $0.276^{*}$ & \fontsize{21}{7}\selectfont $0.109$ & \fontsize{21}{7}\selectfont $0.083$ \\
    
    & \fontsize{21}{7}\selectfont London & \fontsize{21}{7}\selectfont $0.290^{***}$ & \fontsize{21}{7}\selectfont $-0.019$ & \fontsize{21}{7}\selectfont $0.154^{***}$ & \fontsize{21}{7}\selectfont $-0.016^{*}$ & \fontsize{21}{7}\selectfont $-0.041^{***}$ & \fontsize{21}{7}\selectfont $0.010$ & \fontsize{21}{7}\selectfont $0.067^{***}$ & \fontsize{21}{7}\selectfont $-0.037^{**}$ & \fontsize{21}{7}\selectfont $0.044^{***}$ \\
    
    & \fontsize{21}{7}\selectfont Newyork & \fontsize{21}{7}\selectfont $0.233^{***}$ & \fontsize{21}{7}\selectfont $0.193^{***}$ & \fontsize{21}{7}\selectfont $0.087$ & \fontsize{21}{7}\selectfont $0.092^{*}$ & \fontsize{21}{7}\selectfont $0.113^{**}$ & \fontsize{21}{7}\selectfont $-0.015$ & \fontsize{21}{7}\selectfont $0.040^{*}$ & \fontsize{21}{7}\selectfont $-0.024$ & \fontsize{21}{7}\selectfont $-0.093^{***}$ \\
    
    & \fontsize{21}{7}\selectfont Shanghai & \fontsize{21}{7}\selectfont $0.399^{***}$ & \fontsize{21}{7}\selectfont $-0.007$& \fontsize{21}{7}\selectfont $-0.006$ & \fontsize{21}{7}\selectfont $0.013$ & \fontsize{21}{7}\selectfont $-0.042^{***}$  & \fontsize{21}{7}\selectfont $0.038^{**}$ & \fontsize{21}{7}\selectfont $-0.097^{***}$ & \fontsize{21}{7}\selectfont $0.026$ & \fontsize{21}{7}\selectfont $0.005$ \\
    
    & \fontsize{21}{7}\selectfont Beijing & \fontsize{21}{7}\selectfont $0.116^{***}$ & \fontsize{21}{7}\selectfont $-0.161^{***}$ & \fontsize{21}{7}\selectfont $-0.018$ & \fontsize{21}{7}\selectfont $0.051$ & \fontsize{21}{7}\selectfont $0.052^{***}$ & \fontsize{21}{7}\selectfont $0.091^{***}$ & \fontsize{21}{7}\selectfont $0.172^{***}$ & \fontsize{21}{7}\selectfont
    $0.062^{***}$  & \fontsize{21}{7}\selectfont $-0.053^{***}$ \\
    
    & \fontsize{21}{7}\selectfont Guangzhou & \fontsize{21}{7}\selectfont $0.390^{***}$ & \fontsize{21}{7}\selectfont
     $0.102^{***}$ & \fontsize{21}{7}\selectfont $-0.039^{**}$ & \fontsize{21}{7}\selectfont $0.047^{***}$ & \fontsize{21}{7}\selectfont 
     $-0.095^{***}$ & \fontsize{21}{7}\selectfont $0.040$ & \fontsize{21}{7}\selectfont $-0.159^{***}$ & \fontsize{21}{7}\selectfont
       $0.068^{***}$ & \fontsize{21}{7}\selectfont $0.036$ \\
    
    & \fontsize{21}{7}\selectfont Los Angeles & \fontsize{21}{7}\selectfont $0.009$ & \fontsize{21}{7}\selectfont $0.089^{***}$ & \fontsize{21}{7}\selectfont $0.079^{***}$ & \fontsize{21}{7}\selectfont $0.051^{**}$ & \fontsize{21}{7}\selectfont $-0.146^{***}$ & \fontsize{21}{7}\selectfont $-0.068^{***}$ & \fontsize{21}{7}\selectfont $-0.070^{***}$ & \fontsize{21}{7}\selectfont $0.107^{**}$ & \fontsize{21}{7}\selectfont $0.068^{***}$ \\

    \bottomrule
    \end{tabular}
    }    
    \caption{The ATE of different types of buildings on Over illumination, Trespass and Clutter. $^*p< 0.1$; $^{**}p< 0.05$; $^{***}p < 0.01$.}
    \label{result}
\end{table*}

\subsubsection{Identifying Contributors to Light Pollution.}

\begin{figure}[h]
  \centering
  \includegraphics[width=\linewidth]{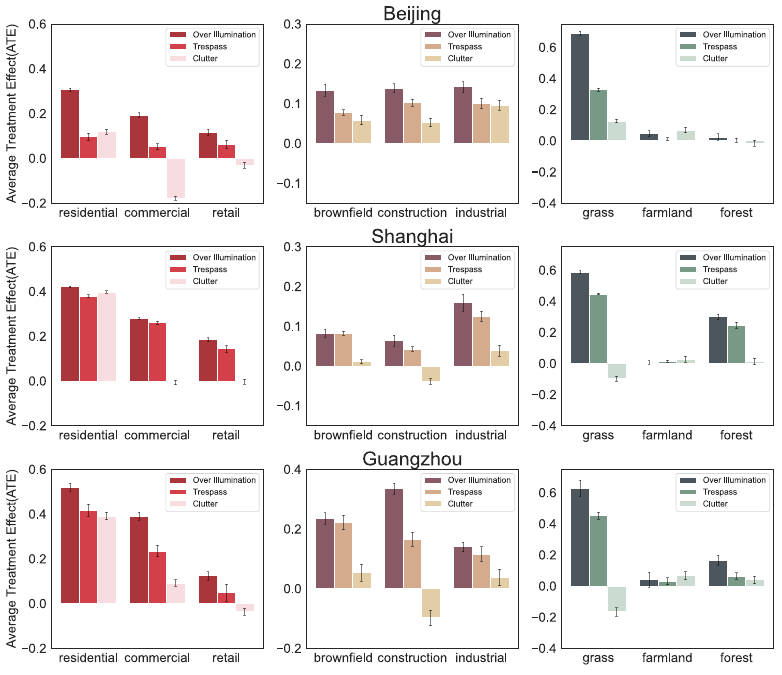}
  \caption{The results of DML in Chinese metropolises.}
  \label{china}
\end{figure}

We conducted comprehensive experiments covering all residential areas across seven metropolises, meticulously documenting the ATE of various building types on light pollution indices. The outcomes are detailed in Table \ref{result}. We categorized the nine building types into three distinct groups: urban living areas (including residential buildings, commercial centers, and retail spaces), urban development areas (encompassing brownfields, construction areas, and industrial zones), and urban landscaping (including grasslands, forests, and farmlands). Comparing their effects allows us to gain valuable insights into light pollution. Due to space constraints, we present results for three Chinese cities only: Beijing, Shanghai, and Guangzhou, as depicted in Figure \ref{china}.

It is evident that the origins of light pollution vary considerably across different metropolises. Notably, \textit{grasslands, commercial centers, and residential buildings have emerged as noteworthy contributors to light pollution}, consistently exhibiting high ATE values across various metropolises.

\subsubsection{Generating Light Pollution Maps.}

\begin{figure}[h]
  \centering
  \includegraphics[width=\linewidth]{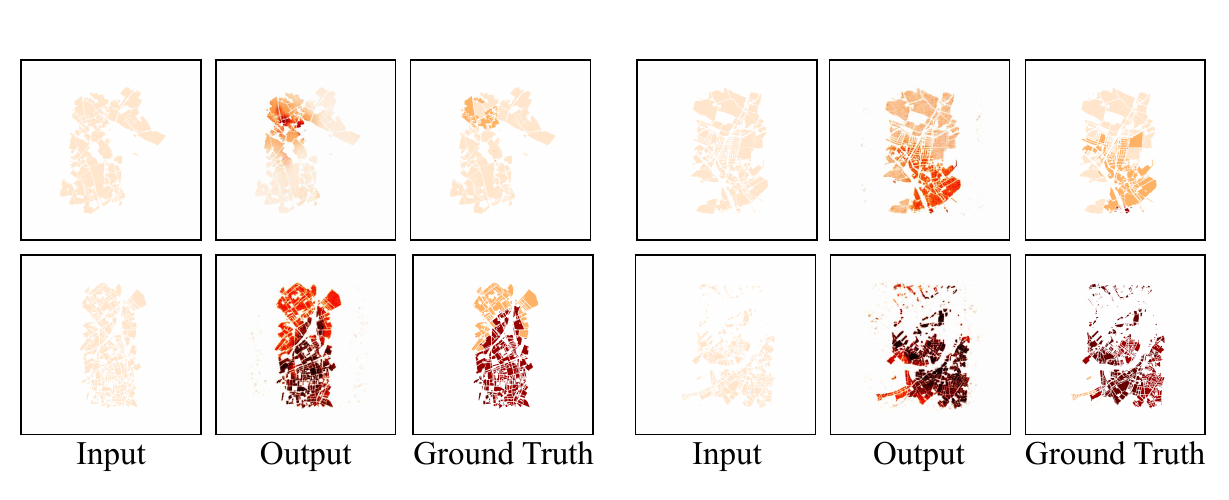}
  \caption{Light pollution maps generated by CAGAN.}
  \label{0123}
\end{figure}

We utilize a well-trained model to generate light pollution maps that facilitate the identification of buildings making notable contributions to light pollution for local residents. Furthermore, the model excels at accurately pinpointing sources of light pollution, thereby enabling a judicious allocation of nighttime lighting resources. Figure \ref{0123} displays the remarkable performance of CAGAN. The model delineates building boundaries within residential areas and strategically leverages conditional information to guide the accurate coloring of land blocks.

\begin{figure}[h]
  \centering
  \includegraphics[width=\linewidth]{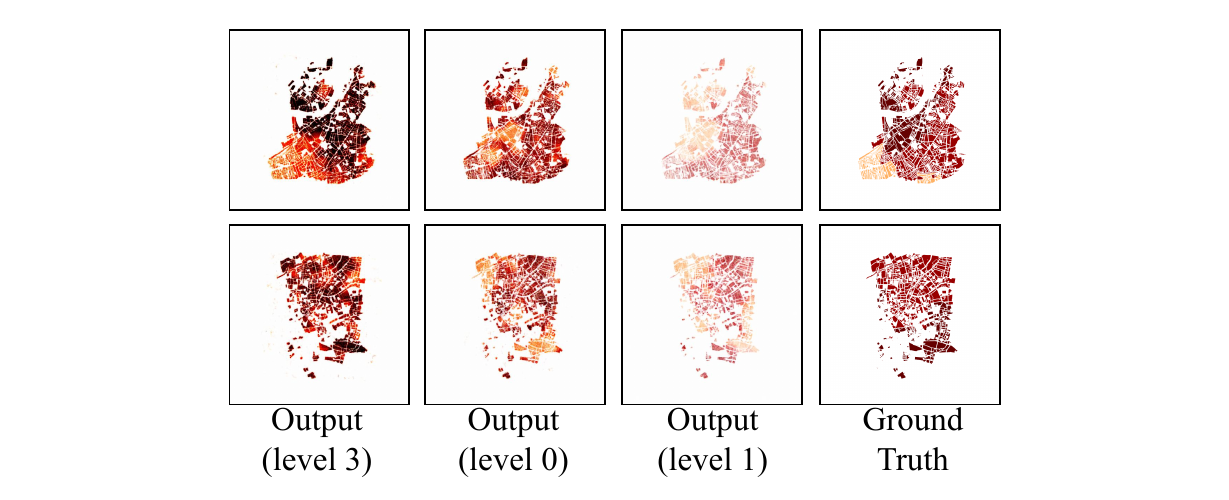}
  \caption{The improvement of severe light pollution.}
  \label{improvement}
\end{figure}

The prowess of CAGAN lies in its capacity to produce light pollution maps grounded in causal relationships that may not correspond to reality. For instance, it can simulate the enhancement of an area previously plagued by intense light pollution. This ability can be attributed to two pivotal factors: firstly, the integration of a well-trained classifier that guides the generator's training, infusing the generated images with both conditional information and interpretability; secondly, the inclusion of a variational inference module that empowers the model to grasp the underlying factors driving light pollution, endowing it with a sense of causality. We have manipulated light pollution levels in residential areas and generated corresponding maps, illustrated in Figure \ref{improvement}.


\subsection{Evaluation of Generated Light Pollution Maps}
We employ a range of metrics to comprehensively evaluate the quality of our generated light pollution maps, spanning from pixel-level analysis to perceptual assessment. These metrics encompass the Inception Score (IS), Peak Signal-to-Noise Ratio (PSNR), Mean Absolute Error (MAE), Multi-Scale Structural Similarity Index Measure (SSIM), Learned Perceptual Image Patch Similarity (LPIPS), and Relative Average Spectral Error (RASE).


\begin{table}[h]
  \centering
  \setlength{\tabcolsep}{3pt} 
    \begin{tabularx}{\linewidth}{XXXXXXX}
    \bottomrule 
     & \multicolumn{2}{|c}{\fontsize{9}{7}\selectfont{ \textbf{Pixel Level}}} & \multicolumn{3}{|c}{ 
\fontsize{9}{7}\selectfont{ \textbf{Perceptual Level}}} \\
     \hline
     \multicolumn{1}{c}{\fontsize{9}{7}\selectfont{ IS $\uparrow$ }} & \multicolumn{1}{|c}{\fontsize{9}{7}\selectfont{ PSNR $\uparrow$ }} & \multicolumn{1}{c}{\fontsize{9}{7}\selectfont{ MAE $\downarrow$ }} & \multicolumn{1}{|c}{\fontsize{9}{7}\selectfont{ SSIM $\uparrow$ }} & \multicolumn{1}{c}{ \fontsize{9}{7}\selectfont{ LPIPS $\downarrow$ }} & \multicolumn{1}{c}{ \fontsize{9}{7}\selectfont{RASE $\downarrow$ }} \\ \hline
     \multicolumn{1}{c}{ \fontsize{9}{7}\selectfont{ 3.253 }} & \multicolumn{1}{|c}{ \fontsize{9}{7}\selectfont{ 24.095 }} & \multicolumn{1}{c}{ \fontsize{9}{7}\selectfont{ 0.012 }} & \multicolumn{1}{|c}{ \fontsize{9}{7}\selectfont{ 0.941 }} & \multicolumn{1}{c}{ \fontsize{9}{7}\selectfont{ 0.068 }} & \multicolumn{1}{c}{ \fontsize{9}{7}\selectfont{ 126.953 }}\\
    \toprule 
    \end{tabularx}%
    \caption{Quality evaluation of CAGAN-generated Maps.}
  \label{IS}%
\end{table}%


Furthermore, we compute the Kullback-Leibler (KL) divergence to compare the probability distributions of pollution levels between the generated and actual light pollution maps, before classifying them using the softmax function. This comparison also extends to random and uniform distributions. A lower KL divergence value here suggests a higher resemblance in pollution levels between the two maps, emphasizing the semantic information of light pollution. The evaluation results are detailed in Tables \ref{IS} and \ref{KL}.

\begin{table}[h]
  \centering
  \begin{tabularx}{\linewidth}{XXXX}
    \bottomrule 
    & \multicolumn{1}{|c}{\fontsize{9}{7}\selectfont CAGAN} & 
    \multicolumn{1}{|c}{\fontsize{9}{7}\selectfont Random} & 
    \multicolumn{1}{|c}{\fontsize{9}{7}\selectfont Uniform} \\
    \hline
    \multicolumn{1}{>{\centering\arraybackslash}X}{\fontsize{9}{7}\selectfont KL $\downarrow$} & 
    \multicolumn{1}{|>{\centering\arraybackslash}X}{\fontsize{9}{7}\selectfont \textbf{7.2380}} & 
    \multicolumn{1}{|>{\centering\arraybackslash}X}{\fontsize{9}{7}\selectfont 36.1757} & 
    \multicolumn{1}{|>{\centering\arraybackslash}X}{\fontsize{9}{7}\selectfont 28.2567} \\
    \toprule 
  \end{tabularx}
  \caption{KL Divergence Comparison}
  \label{KL}
\end{table}


As depicted in Tables \ref{IS} and \ref{KL}, our Causally Aware Generative Adversarial Network (CAGAN) exhibits outstanding performance across both pixel-level and perceptual-level metrics, affirming the practicality of the generated light pollution images. Moreover, the generated light pollution maps successfully capture the probability distribution of light pollution levels. Notably, when compared to uniform and random distributions, our model demonstrates a lower Kullback-Leibler divergence, emphasizing its fidelity to the probability distribution observed in real images.



\subsection{Discussions}
\subsubsection{Causality Interpretability.}

\begin{figure}[h]
  \centering
  \includegraphics[width=\linewidth]{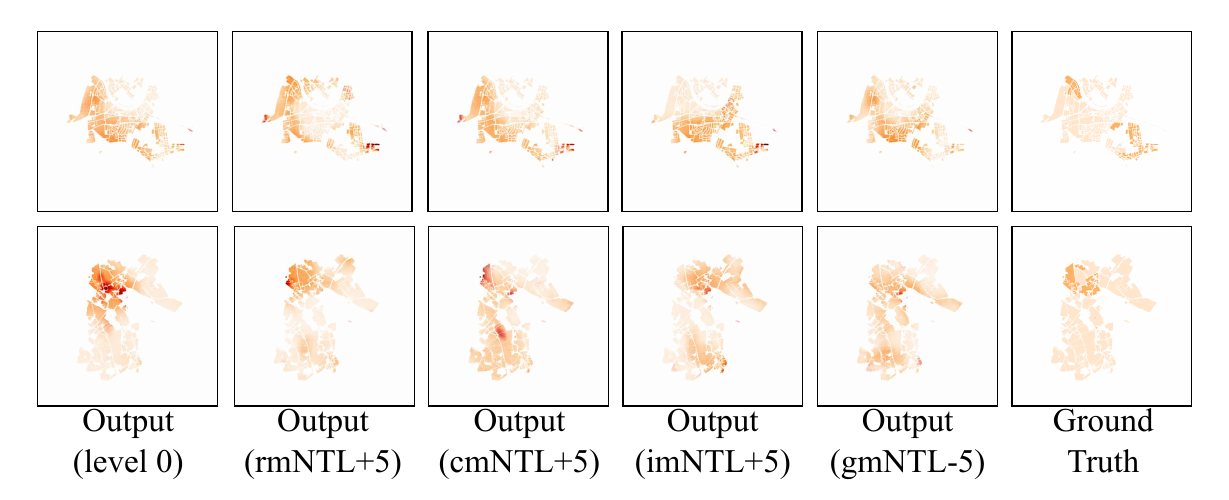}
  \caption{The residential areas after altering the Average NTL (Nighttime Light Intensity) of different buildings.}
  \label{change}
\end{figure}

The generated images exhibit inherent causal interpretability, facilitating a deeper understanding of the outcomes from causal experiments. For instance, deliberately adjusting the average nighttime light intensity of various building types within residential areas leads to observable changes in brightness across different regions. By adjusting the NTL values for Residential, Commercial, Construction, and Grassland categories and subsequently regenerating the light pollution maps, as illustrated in Figure \ref{change}, our model showcases its capacity to assist local administrators in making more informed decisions regarding the allocation of nighttime lighting resources.

\subsubsection{Further Insights.}
Considering both the generated images and causal outcomes in a holistic manner allows us to elucidate and dissect intricate light pollution phenomena. For example, commercial centers often exhibit continuous nighttime illumination, encompassing buildings, billboards, and lighting clutter. Additionally, in economically developed metropolises, densely populated residential zones often experience mutual light trespass. 
Furthermore, alterations in NTL values within grasslands yield broader effects. Urban landscaping areas like grasslands require nighttime illumination for visual enhancement. This necessity might lead to increased lighting installations around such spaces, thus contributing to elevated light pollution. Additionally, grassy areas commonly lack tall buildings and trees, rendering the scattering and reflection of light in the surrounding environment more pronounced, thereby influencing a larger area.

\section{Conclusion}

In a nutshell, the proposed Causally Aware Generative Adversarial Network (CAGAN) introduces a novel approach to enhance interpretability and formulate management strategies for mitigating light pollution in urban residential areas. Through the exploration of causal relationships associated with light pollution in our framework, we have gained a deeper understanding of this phenomenon, paving the way for the implementation of effective mitigation strategies. The investigation has demonstrated the efficacy of CAGAN in assessing residential light pollution across seven economically advanced metropolises. Looking forward, this method holds the potential for broader applications, covering different categories and levels of economic development.

\section*{Acknowledgement}

This work was supported by the National Natural Science Foundation of China (NSFC Grant No.62106274); the Fundamental Research Funds for the Central Universities, Renmin University of China (No.22XNKJ24). We also wish to acknowledge the support provided by the Intelligent Social Governance Platform, Major Innovation \& Planning Interdisciplinary Platform for the "Double-First Class" Initiative.

\bibliography{aaai24}

\end{document}